\documentclass[aps,preprint,prd,showpacs,nofootinbib]{revtex4}

\usepackage{amsmath}
\usepackage{graphicx}
\usepackage{subfigure}
\usepackage{bm}
\usepackage{amssymb}
\usepackage{mathtools}
\usepackage{enumerate}
\usepackage{color}
\usepackage[colorlinks,linkcolor=magenta,anchorcolor=cyan,citecolor=blue]{hyperref}

\newcommand{\ndp}{\left(\frac{\dot{\phi}}{M_p\Lambda}\right)}

\newcommand{\hn}{\left(\frac{H}{\Lambda}\right)}
\newcommand{\don}{\mathcal{D}_1}
\newcommand{\dth}{\mathcal{D}_3}
\newcommand{\cs}{c_s^2}
\newcommand{\A}{\mathcal{A}}

\begin{document}

\title{Positivity in the effective field theory of cosmological perturbations}

\author{Gen Ye$^{1}$\footnote{yegen14@mails.ucas.ac.cn}}
\author{Yun-Song Piao$^{1,2}$\footnote{yspiao@ucas.ac.cn}}

\affiliation{$^1$ School of Physics, University of Chinese Academy of
    Sciences, Beijing 100049, China}

\affiliation{$^2$ Institute of Theoretical Physics, Chinese
    Academy of Sciences, P.O. Box 2735, Beijing 100190, China}

\begin{abstract}

Requiring the existence of a unitary, causal and local
UV-completion places a set of positivity bounds on the
corresponding effective field theories (EFTs). We discuss the obstructions and possibility in applying the positivity bound to cosmology, in particular the EFT of cosmological perturbations.
Taking a $c_T=1$ beyond-Horndeski EFT as an illustrative example,
we derive such bounds, which incorporate the cosmological
correction of order $H^2/\Lambda^2$, $\Lambda$ being the
cutoff scale. The derived bounds are applied to slow-roll inflation with beyond Horndeski operators. It is found that the cosmological positivity bounds may be either stronger or weaker than their flat space counterpart.

\end{abstract}
\maketitle
\section{Introduction}
The effective field theory (EFT) of cosmological perturbations is a
powerful tool to study perturbations around a given cosmological
background. Since the EFT of inflation \cite{Cheung:2007st}, the
relevant idea has been also applied to other cosmological fields,
such as the EFT of dark energy
\cite{Gubitosi:2012hu,Bloomfield:2012ff,Gleyzes:2013ooa,Langlois:2017mxy}
which captures the physics of scalar-tensor theories
\cite{Horndeski:1974wa,Deffayet:2011gz,Kobayashi:2011nu,Gleyzes:2014dya,Langlois:2015cwa}
(for a review, see \cite{Langlois:2018dxi,Kobayashi:2019hrl}) at
the cosmological scale. As another example, based on the EFT
approach, it has been found that fully stable nonsingular
cosmologies exist in theories beyond Horndeski
\cite{Cai:2016thi,Creminelli:2016zwa,Cai:2017dyi,Cai:2017tku,Kolevatov:2017voe,Mironov:2018oec,Ye:2019frg,Ye:2019sth}.

The full UV-complete theory of gravity is yet unknown. Instead of
starting top-down from a UV theory, usually one works directly
with the EFT, which captures the physics of the underlying
theory at certain scales. However, not all low-energy EFTs have a consistent UV
theory, see e.g.\cite{Ooguri:2006in,Obied:2018sgi}.
Assuming the UV-complete theory is causal, unitary and local
Lorentz-invariant, one can derive dispersion relations relating
the IR limit of a scattering amplitude with its UV behavior, which
place a set of bounds on the properties of the corresponding
low-energy EFTs, the so-called positivity bound
\cite{Adams:2006sv,Nicolis:2009qm,Bellazzini:2014waa,Bellazzini:2016xrt,deRham:2017avq,deRham:2017zjm,Chandrasekaran:2018qmx,deRham:2018qqo,Tokuda:2019nqb}.
The positivity bounds have been applied to various EFTs
\cite{Bellazzini:2015cra,Cheung:2016yqr,Cheung:2016wjt,deRham:2017imi,Bellazzini:2017fep,deRham:2017xox,Bellazzini:2019bzh,Bellazzini:2019xts}.


Recently, Ref.\cite{Melville:2019wyy} derived positivity
bounds on a covariant shift-symmetric Horndeski theory (which
might explain the current accelerated expansion), and paired these
bounds with a cosmological parameter estimation analysis. It is
natural to ask what will happen if we incorporate the
cosmological background evolution. It is convenient for our
purpose to work with the EFT of cosmological
perturbations. However, the background evolution is itself the
biggest obstruction in applying the positivity arguments because
it breaks time-translation symmetry and makes Lorentz-invariant
scattering ill-defined. In the limit of a Lorentz-invariant
background, the positivity bounds saturated by the UV completions
of single-field inflation have been investigated in
\cite{Baumann:2015nta}.

Theoretically, an EFT is obtained after one integrates out physics
above the cutoff scale $\Lambda$. On the other hand, the curvature
of a homogeneous FRW universe is proportional to $H^2, \dot{H}$,
so one can use $H^2, \dot{H}$ as additional dimensionful
parameters in constructing the EFT  \cite{Kaloper:2002uj,Afkhami-Jeddi:2018own,Afkhami-Jeddi:2018apj}. In
this paper, we assume that the background varies slowly enough so that we can treat the
$\mathcal{O}(H^{2n}/\Lambda^{2n},\dot{H}^n/\Lambda^{2n})$
corrected Lagrangian coefficients as constants \cite{Baumann:2015nta,Kaloper:2002uj,Arkani-Hamed:2015bza,Arkani-Hamed:2018kmz} in calculation of the scattering amplitude. Despite these corrections, positivity argument in cosmology is still plagued by difficulties such as asymptotic states and particle production in the cosmological background, which are expected to introduce unknown corrections scale as $\mathcal{O}(H^2/M^2)$, $M$ being the scattering energy scale, in the bound itself. We will argue in section-\ref{subsec:positivity in cosmology} that under certain assumptions it might be possible to extract part of the positivity information in cosmology even if one does not know how to calculate the $\mathcal{O}(H^2/M^2)$ corrections in the positivity bound. We then derive the bounds, incorporating the cosmological background, for a specific $c_T=1$ beyond-Horndeski EFT.


This paper is structured as follows. The effective Goldstone
Lagrangian is derived in section-\ref{sec:EFT}. Section-\ref{subsec:positivity argument} presents the familiar Minkowski positivity bound. Section-\ref{subsec:positivity in cosmology} discusses existing obstructions in deriving a cosmological positivity bound, what corrections they may introduce and under what circumstance can we neglect them. Then in section-\ref{subsec:apply bound}, we obtain bounds with leading cosmological
correction. Applications of our bounds to the cosmolgical
scenarios of interest are discussed in
section-\ref{subsec:example}.

\section{The effective Goldstone Lagrangian}\label{sec:EFT}

We first derive the EFT of cosmological
perturbations to be bounded by positivity. We want to focus on the Goldstone EFT, so we will
work in the decoupling limit \cite{Cheung:2007st}. Such limit,
however, becomes subtle in a $c_T\ne1$ theory
\cite{Creminelli:2014wna}. It is also noticed that the EFTs of
modified gravity at cosmological scales have been strictly
constrained by GW170817 to $c_T=1$
\cite{Abbott:2018lct,Creminelli:2017sry,Sakstein:2017xjx,Langlois:2017dyl} (note that it may not be the case at scales not observed by LIGO \cite{deRham:2018red}). Also,
as mentioned, the stable nonsingular cosmological models can be
implemented only in theories beyond Horndeski (We have little evidence whether $c_T=1$ in the early universe though.). Thus we are well-motivated to consider a $c_T=1$ beyond-Horndeski EFT as
an illustrative example\footnote{The theory may receive other strict constraints if one considers stability of gravitational waves \cite{Creminelli:2018xsv,Creminelli:2019nok,Creminelli:2019kjy}.}.

The shift-symmetric $c_T=1$ beyond-Horndeski theory can be written
as \cite{Gleyzes:2014dya,Creminelli:2017sry}
\begin{equation}\label{L_bH}
L=M_p^2\Lambda^2\left[B(X)\frac{R}{\Lambda^2}+G_2(X)+G_{3}(X)\frac{\Box\phi}{M_p\Lambda^2}-\frac{4}{X}B_X\frac{\phi^\mu\phi^\nu\phi_{\mu\nu}\Box\phi-\phi^\mu\phi_{\mu\nu}\phi_\lambda\phi^{\lambda\nu}}{M_p^4\Lambda^6}\right],
\end{equation}
where $M_p$ is the reduced Planck mass and $\Lambda$ is the EFT
cutoff\footnote{In self-accelerating cosmologies one often takes $\Lambda\sim H$ in the Galileon terms to have an order one effect on the background. We do not consider such models in this paper and will always assume $\Lambda\gg H$.}. All coefficients $B(X)$, $G_{2,3}(X)$ and fields are
dedimensionalised by $\phi\to \phi/M_p, \ \partial\to
\partial/\Lambda$ and $X\equiv\phi^\mu\phi_\mu/(M_p^2\Lambda^2)$.
Subscript $X$ denotes partial derivatives with respect to $X$, for
instance $B_X\equiv \frac{\partial}{\partial X}B$. 

In the unitary gauge ($\delta \phi=0$), the Lagrangian
\eqref{L_bH} is equivalent to \cite{Creminelli:2017sry}
\begin{equation}\label{unitary L}
L=M_p^2\Lambda^2\left[G_2(X)+Q(X)\frac{K}{\Lambda}+B(X)\frac{R^{(3)}+K^\mu_{\
\nu}K^{\nu}_{\ \mu}-K^2}{\Lambda^2}\right]
\end{equation}
where $K^{\nu}_{ \ \mu}$ and $R^{(3)}$ are the extrinsic curvature tensor and Ricci
scalar of the spacelike uniform-$\phi$ hypersurface, respectively, and $Q(X)\equiv
-\int \sqrt{-X}G_{3X}(X)dX$.

An arbitrary time slicing $(t,\bm{x})$ is related to the unitary
gauge time $(\tilde{t},\tilde{\bm{x}})$ by
$\tilde{t}=t+\pi(t,\bm{x})$ and $\bm{x}=\tilde{\bm{x}}$, then
\begin{equation}\label{goldstone-metric}
\tilde{g}^{00}(\tilde{t},\tilde{\bm{x}})=\frac{\partial \tilde{t}}{\partial x^\mu}\frac{\partial \tilde{t}}{\partial x^\nu}g^{\mu\nu}(t,\bm{x})=(1+\dot{\pi})^2g^{00}+2(1+\dot{\pi})\partial_i \pi g^{0i}+g^{ij}\partial_i \pi \partial_j \pi,
\end{equation}
where $\pi$ is a Goldstone field. Quantities in the unitary
gauge are labeled with tildes.
Due to the derivative
coupling, the scattering process $\pi\pi\to\pi\pi$ is dominated by
sub-Hubble contribution, where the Goldstone mode decouples from
gravity \cite{Cheung:2007st}. We are thus allowed to choose the
standard FRW metric
\begin{equation}\label{FRW}
ds^2=g_{\mu\nu}dx^\mu dx^\nu=-dt^2+a^2(t)\delta_{ij}dx^idx^j
\end{equation}
as the metric in the new coordinate system $(t,\bm{x})$. Relevant
Stuckelberg tricks are given in
Appendix-\ref{apdx:Stuekelberg}.





It is convenient to define the dimensionless parameters
\begin{equation}\label{epsilon}
\epsilon\equiv\frac{\dot{\phi}^2/2}{M_p^2H^2},\qquad\epsilon_H\equiv-\frac{\dot{H}}{H^2},
\end{equation}
where $\epsilon_H$ describes evolution of the universe. We will
assume $|\ddot{\phi}/(H\dot{\phi})|\ll1$, so that we can safely neglect $\ddot{\phi}$. 
In a $\phi$-dominating universe,
$\epsilon$ is determined by the Friedman equation and related to $\epsilon_H$ by
the equation of motion of $\phi$, in particular,
$\epsilon=\epsilon_H$ if $\phi$ is canonical.
We thus have the time derivative of a function $f$ as
\begin{equation}\label{int by part}
\frac{df}{dt}=\left(-\epsilon_H H^2\frac{\partial}{\partial H}+\epsilon^{(1)}_H\frac{\partial}{\partial \epsilon_H}\right)f,
\end{equation}
where $\dot{\epsilon}_H\equiv H\epsilon^{(1)}_H\simeq2H\epsilon_H^2$ if $\phi$ dominates.

At the tree level, it is sufficient to consider $S^{(2)}$, $S^{(3)}$ and $S^{(4)}$. We assume $S^{(2)}$ dominates and $S^{(3)}$, $S^{(4)}$ are under perturbative control \cite{Ashoorioon:2018uey}. Exploiting the fact that the scattering process is almost instant compared to the variation of the background field $\phi$ (i.e: $\bar{X}$), after some integration by parts, one has
\begin{equation}\label{L2}
S^{(2)}=(M_p\Lambda)^2\int dx^4\sqrt{-g}\left[U\dot{\pi}^2-V\frac{(\partial_i\pi)^2}{a^2}\right],
\end{equation}
where
\begin{subequations}
\begin{equation}
U=\bar{X}\bar{G}_{2X}-6\bar{X}\hn^2\bar{B}_X-6\hn(-\bar{X})^{3/2}\bar{G}_{3X}+2\bar{X}^2\bar{G}_{2XX}+\dots,
\end{equation}
\begin{equation}
V=\bar{X}\bar{G}_{2X}-4(7-4\epsilon_H)\bar{X}\hn^2\bar{B}_X-4(-\bar{X})^{3/2}\hn \bar{G}_{3X}+\dots.
\end{equation}
\end{subequations}
All bared quantities are evaluated on the
background
$\bar{X}=-\dot{\phi}^2/(M_p^2\Lambda^2)=-2(H/\Lambda)^2\epsilon$.
In the rest of this paper, the results are presented
using $\epsilon$ and $\epsilon_H$ instead of $\bar{X}$ and
$\dot{H}$. We will also suppress the upper bar of background quantities for simplicity. The sound speed squared is
\begin{equation}\label{sound speed}
c_s^2\equiv
V/U=1+\frac{2}{G_{2X}}\left(4(-1+\epsilon_H)B_X+\sqrt{2}
\epsilon^{1/2} G_{3X}+2 \epsilon G_{2XX}\right)\hn^2+\dots,
\end{equation}
which is approximately constant since
$\frac{d}{dt}c_s^2/\Lambda\sim\epsilon_H H^3/\Lambda^3$ is of
higher order in $H/\Lambda$. Truncation to leading order in $H/\Lambda$ is valid if
the correction is $<{\cal O}(1)$\footnote{Now we have $\bar{X}=-\dot{\phi}^2/(M_p^2\Lambda^2)\sim const.$ and $H/\Lambda<{\cal O}(1)$. To guarantee a slowly varying background, one should, in principle, additionally require $-\epsilon_H H/\Lambda\ll1$. This condition is automatically satisfied for physical value of $\epsilon_H$ when $H/\Lambda\ll1$ is true.}. For unity $B_X,\epsilon
G_{2XX},\epsilon^{1/2}G_{3X}$, it is sufficient to have
$H\lesssim0.1\Lambda$. Rescale the spacial coordinates $x^i\to
a c_s x^i$ and define the canonically normalised field
$\pi_c=\sqrt{2c_s^3U}M_p \Lambda\pi$, we have
\begin{equation}\label{quadratic Lagragian}
L^{(2)}=\frac{1}{2}(\partial\pi_c)^2,
\end{equation}
where $(\partial\pi_c)^2=\dot{\pi}_c^2-(\partial_i\pi_c)^2$.

The final Lagrangian for the UV scattering is (the subscript $c$
is omitted)
\begin{equation}\label{scattering Lagrangian}
\begin{aligned}
L=&\frac{1}{2}(\partial\pi)^2+\frac{1}{M_p\Lambda}\left[\alpha_1\dot{\pi}^3+\alpha_2\dot{\pi}(\partial\pi)^2\right]+\frac{1}{M_p^2\Lambda^2}\Big[\beta_1\dot{\pi}^4+\beta_2\dot{\pi}^2(\partial\pi)^2+\beta_3[(\partial\pi)^2]^2\Big]\\&+\frac{\beta_4}{M_p^2\Lambda^4}
\left[\dot{\pi} \partial_i\partial_j\pi \partial_i\dot{\pi}
\partial_j \pi-\dot{\pi}\partial_i\dot{\pi}\partial_i \pi
\nabla^2\pi \right],
\end{aligned}
\end{equation}
see Appendix-\ref{apdx:L calculation} for the details of
derivation\footnote{The Lagrangian \eqref{scattering Lagrangian} is non-relativistic. Non-relativistic vertices with odd number of derivatives may contribute amplitude proportional to $s^{n/2}$, resulting in new branch cuts in the $s$ plane, see Ref.\cite{Baumann:2015nta}. As is shown in Appendix-\ref{apdx:L calculation}, fortunately, higher derivative operators are under control in the theory we consider and no new branch cuts are generated at the order of interest. Also \eqref{scattering Lagrangian} does not contribute any other non-analyticities more complicated than the familiar cuts and poles crossing symmetric in the complex $s$ plane.}. The explicit expressions of $\alpha,\beta$ in terms of
$G_2(X)$, $G_3(X)$ and $B(X)$ in \eqref{L_bH} are given in
Appendix-\ref{apdx:coefficient}. The coefficients of the final
Lagrangian \eqref{scattering Lagrangian}, as expected, do not
contain $B(X)$ or $Q(X)$, since in the covariant Lagrangian \eqref{L_bH} the Ricci
curvature $R$ is invariant and only the $X$ dependent part of
$G_3(X)$ is relevant, though they appear in coefficients before
individual vertices (for example, eq.\eqref{S4 modification} contains
$B(X)$). This can serve as a quick consistency check of calculation.

\section{Positivity bounds in cosmology}\label{sec:bound with background}

\subsection{Positivity in flat space}\label{subsec:positivity argument}

We briefly review the Minkowski positivity argument, see
e.g.\cite{Adams:2006sv,Bellazzini:2017fep,deRham:2017avq} for details. In a Lorentz-invariant UV
theory with unitarity, locality and causality, the $2\to2$ scattering
amplitude is expected to be an analytic function of the
Mandelstam variables $(s,t,u)$ with poles and branch cuts, satisfying the Froissart-Martin bound 
\cite{Froissart:1961ux,Martin:1962rt}.

\begin{figure}
	\includegraphics[width=5in]{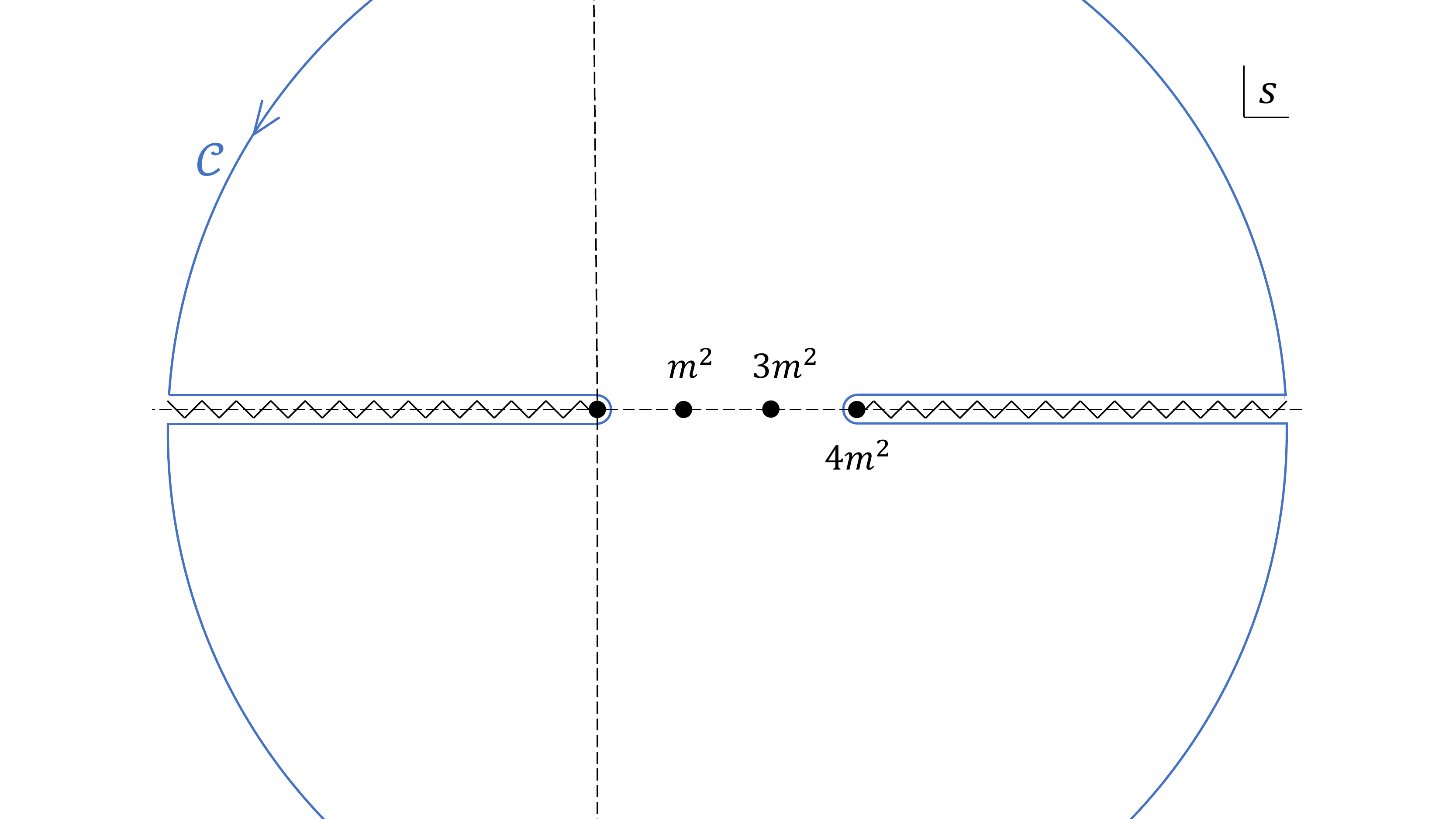}
	\caption{The analytic structure of $\A(s)$ in the forward limit
		and the integration contour $\mathcal{C}$. The branch cut starts
		from the biparticle production threshold and extends all the way
		to infinity.}
	\label{analytic structure}
\end{figure}

Consider a massive scalar field with mass $m$. The Mandelstam
variables are not independent $s+t+u=4m^2$. We denote the $2\to2$
amplitude by $\A(s,t)$. For fixed $t$, it is an analytic function
$\A_t(s)$ of $s$. $\A_t(s)$ can be
extended to the complex s-plane by crossing symmetry and analytic continuation. Consider the following Cauchy integral at
fixed $t$
\begin{equation}\label{Cauchy int}
\sum Res\left(\frac{\A_t(s)}{(s-\mu^2)^3}\right)=\frac{1}{2\pi i}\oint_\mathcal{C} \frac{\A_t(s)}{(s-\mu^2)^3} ds.
\end{equation}
Exactly speaking, the LHS of \eqref{Cauchy int} should be an integral over all non-analyticities enclosed by $\mathcal{C}$. However, since the analytic structure of \eqref{scattering Lagrangian} is simple, we explicitly write down the integral as sum of residues. Typical positivity argument then requires pushing to the forward limit ($t\to0$). However, such a limit is singular in a theory with massless gauge particles, the gravitational theory in particular (see e.g.\cite{Adams:2006sv}). Recently, it is suggested that such singularity can be regulated by compatifying one spatial dimension \cite{Bellazzini:2019xts}. The resulting amplitude (with the Kaluza-Klein modes substracted) remains finite in the forward limit but receives additional contribution from the dilaton and graviphoton generated by compactification. Fortunately, in the theory \eqref{L_bH} considered here, the dilaton and graviphoton do not contribute at the order  of interest. Therefore, in effect, we can simply drop the singular term $\sim\frac{s^2}{M_p^2t}$ and continue with the standard positivity argument. Now we are allowed to pass to the
forward limit $t\to0$. Subscript $t$ will be dropped whenever we are in the forward limit. The analytic structure of $\A(s)$ and the
integration contour $\mathcal{C}$ are depicted in
Fig.\ref{analytic structure}. By the Froissart-Martin bound
$|\A_t(s)|\sim \mathcal{O}(s\ln^2s)$ as $s\to\infty$
\cite{Froissart:1961ux,Martin:1962rt}, the contribution from the semi-circular arcs of the contour go to zero when pushed to
infinity. We are thus left with $\frac{1}{\pi}\left(\int_{-\infty}^{0}+\int_{4m^2}^{+\infty}\right)\frac{Im(\A_t(s))}{(s-\mu^2)^3}ds$. By the optical theorem we arrive at the following equality
\begin{equation}\label{branch int}
\frac{1}{2\pi i}\oint_\mathcal{C} \frac{\A(s)}{(s-\mu^2)^3} ds=\frac{1}{\pi}\int_{4m^2}^{+\infty}\sqrt{1-\frac{4m^2}{s}}\left(\frac{s\sigma^{2\to any}(s)}{(s-\mu^2)^3}+\frac{s\sigma^{2\to any}(s)}{(s-4m^2+\mu^2)^3}\right)ds.
\end{equation}
RHS is positive definite if $0<\mu^2<4m^2$. There are three poles included on LHS. Since the EFT \eqref{scattering Lagrangian} only contains derivative
interactions, residues at $s=m^2$ and $s=3m^2$, associated
with propagators in the exchange diagrams, are proportional to
powers of $m^2$ and thus become negligible when $s\gg m^2$. Then the final
bound is
\begin{equation}\label{the bound}
\A''(\mu^2)\ge0.
\end{equation}
Since $s\gg m^2\sim\mu^2$, this is equivalently $\A''(s\to0)\ge0$. Besides, combine
the optical theorem with the partial wave expansion
$\A(s,t)=16\pi\sqrt{s/(s-4m^2)}\sum_l
(2l+1)P_l(\cos\theta)a_l(s)$, one obtains another bound for
$s\ge4m^2$ \cite{Martin:1965jj} (for recent exploitation of this
bound, see e.g.\cite{deRham:2017avq})
\begin{equation}\label{the t bound}
\frac{\partial^n}{\partial t^n}Im[\A(s+i\epsilon,0)]\Big|_{t=0}\ge0.
\end{equation}
Apply this bound to Eq.\eqref{Cauchy int}, one reaches the positivity of
$\partial_t^n\A''(s\to0,t\to 0)$.

\subsection{Positivity in cosmology} \label{subsec:positivity in cosmology}
Conventional EFTs (as
well as the underlying UV theories) are Poincare invariant. The
EFT of cosmological perturbations, on the other hand, are obtained
by breaking the time-diffeomorphism invariance with a gauge choice
(unitary gauge). When the time-translation symmetry is broken
because of the evolving background, the notion of Mandelstam
variables becomes ill-defined. We focus on scattering process with energy scale $M^2\gg m^2$ because the Goldstones are derivatively coupled, while keeping $M^2\ll\Lambda^2$ so the EFT remains valid.

The cosmological background evolution manifests in the study of positivity mainly in two ways. One is the cosmological correction in the Lagrangian coefficients, which we have already considered in the previous section. In a slowly varying background, we will treat these Lagrangian coefficients as constants when calculating the amplitude \cite{Baumann:2015nta,Arkani-Hamed:2015bza,Arkani-Hamed:2018kmz}. Another is that the time-dependent background affects the perturbative calculation of the amplitude and breaks time translation symmetry. These effects will further influence the positivity derivation. We do not yet know how to obtain a robust positivity bound in an evolving background. It has to be pointed out that we will not derive a reliable bound in some general time-dependent curved background in this subsection. Instead, we will analyze the major obstructions in obtaining such a bound and argue that when certain assumptions hold, one might use \eqref{the bound} and \eqref{the t bound}, neglecting cosmological corrections to the bound, to extract some of the positivity information in cosmology.

We start from the cosmological mode function
\begin{equation}
u_k(\tau)=\sqrt{\frac{\pi}{4}}\sqrt{\tau}H^{(1,2)}_\nu(|k\tau|).
\end{equation} 
This mode function assumes flat space solution (up to a constant phase shift) $\frac{1}{\sqrt{2k}}e^{-ik\tau}$ on scales much smaller than the curvature ($|k\tau|\gg1$). For the universe with $a\propto t^p$ ($p=1/2$ if radiation dominant, $p=2/3$ if matter dominant), $\nu=\frac{|3p-1|}{2|-p+1|}$, while for quasi-de Siiter space $\nu\simeq 3/2$. For inflationary universe $\tau\in(-\infty,0)$ so modes are initially sub-Hubble, while for decelerated expansion $\tau\in(0,+\infty)$ so modes finally enter horizon. Due to the $e^{\pm ik\tau}$ asymptotic behavior of $u_k(\tau)$ and the relation $H^{(1)}_\nu(ze^{i\pi})=-e^{-i\nu\pi}H^{(2)}_\nu(z)$ one can obtain the free field Feynman propagator in the usual way by calculating free field two point function\footnote{There is possible divergence at $k^0=0$ since $H_\nu^{(1,2)}(z\to0)\sim z^{-\nu}$. This is regulated by the negligible scalar mass $m$ so that $H^{(1,2)}_\nu(k\tau)\to H^{(1,2)}_\nu(\omega\tau)$ with $\omega^2=k^2+m^2>0$, thus the integration does not encounter any divergence at $k^0=0$}
\begin{equation}
D_F(x,y)=\int\frac{dk^4}{(2\pi)^4}\frac{ik^0}{k^2+i\epsilon}\frac{\pi}{2}\sqrt{x^0y^0}H^{(2)}_\nu(k^0x^0)H^{(1)}_\nu(k^0y^0)e^{i\bm{k}\cdot(\bm{x}-\bm{y})}.
\end{equation}
The propagator depends on $x$ and $y$ separately since time translation is broken. Using the asymptotic expansion of Hankel functions the leading irreversible part of $D_F(x,y)$ is estimated to be of order $\frac{x^0-y^0}{kx^0y^0}\sim\frac{1}{k^2x^0y^0}$.

The first obstruction in evaluating the amplitude is the construction of asymptotic state $|\Omega\rangle$. Due to background evolution, the energy argument used in Minkowski to obtain $|\Omega\rangle\propto \exp\left(-iT\int_{-\infty}^{t_0} H_I dt\right)|0\rangle$ is no longer applicable. Moreover, the amplitude $\langle n_1 \dots n_i|\Omega\rangle$, $\langle n_1 \dots n_i|$ being occupation states at some later time, is generally non-vanishing because of the cosmological evolution. However, since we only study sub-Hubble scattering process, we assume an asymptotic state parallel to the Minkowski one
\begin{equation}
|\Omega\rangle\propto \exp\left(-iT\int_{\tau_i}^{\tau_0} H_I d\tau\right)|0\rangle,
\end{equation}
where $\tau_0$ is the time of scattering and $\tau_i$ is some time well after the scattered momentum modes entering horizon\footnote{In the rest of this subsection we assume that the universe is in decelerated expansion and modes enter horizon. The calculation is the same for accelerated expansion with the substitution $\tau\to-\tau$}. By this assumption we have neglected all particle production effects (and in particular any non-perturbative non-analyticities they may bring about) from the time-dependent background. It is only valid for sub-Hubble high energy scattering, a more careful treatment of asymptotic states in de Sitter spacetime is presented in \cite{Marolf:2012kh}. Now the $2\to2$ amplitude is $\langle p_1 p_2|\exp\left(-iT\int_{\tau_i}^{+\infty} H_I d\tau\right)|p_3 p_4\rangle$. The external legs are sub-Hubble with definite momentum $\bm{p}_i$, we thus further assume contraction of all external legs yields the flat space solution $e^{-ip\cdot x}$. We should, however, keep the cosmological mode function $H_{\nu}$ in the propagator since the $k^0$ integral crosses $k^0=0$. This assumption is crucial in preserving crossing symmetry $\omega\leftrightarrow -\omega$, $\bm{k}\leftrightarrow-\bm{k}$ at the Feynman diagram level.

With assumptions discussed above, the only difference in evaluating the tree level $2\to2$ amplitude is the integral $\int dx^0$ and $\int dy^0$. In Minkowski, these two enforce energy conservation between the initial and final states i.e: $\delta(p_1+p_2-k^0)\delta(k^0-p_3-p_4)$. In an evolving background, we have to estimate
\begin{equation}\label{k0 int}
\begin{aligned}
I&=\int_{\tau_i}^{+\infty}dx^0 e^{i(p_1+p_2)x^0}\sqrt{\frac{\pi}{2}}\sqrt{k^0x^0}H^{(2)}_{\nu}(k^0x^0)\sim \int_{\tau_i}^{+\infty}dx^0 e^{i(p_1+p_2-k^0)x^0}(1- \frac{ia(\nu)}{k^0x^0})\\&\sim\frac{ie^{i(p_1+p_2-k^0)\tau_i}}{p_1+p_2-k^0}-\frac{ia(\nu)}{k^0}\Gamma(0,-i|p_1+p_2-k^0|\tau_i)
\end{aligned}
\end{equation}
where $a(\nu)=(4\nu^2-1)/8$. The last line is obtained by slightly rotating the integration line in the complex plane (i.e. $\infty\to(1\pm i\epsilon)\infty$). The integrand of $\int dk^0/(2\pi)$ is regular (note that the propagator $\frac{i}{k^2+i\epsilon}$ has no pole on the real line due to $i\epsilon$) so the first term in \eqref{k0 int} is simply $-2\pi i\delta(p_1+p_2-k^0)$, which enforces energy conservation. The correction term is characterized by an incomplete gamma function, which peaks at $k^0=p_1+p_2$ and oscillates and decays as $|p_1+p_2-k^0|\to \infty$ with $\int_0^{+\infty}\Gamma(0,-ix)dx=i$ and $\lim_{\epsilon\to0}\int_{-\epsilon}^{+\epsilon}\Gamma(0,-ix)dx=0$. In particular it is the correction part that contributes when $p_1+p_2\ne k^0$. Based on these assumptions and observations we propose
\begin{equation}\label{assertion}
i\A(p_1p_2\to p_3p_4)=i\A_{min}\delta^{(4)}(\Sigma p)+i\A_{cos}\rho(\Delta E)\delta^{(3)}(\Sigma \bm{p}).
\end{equation}
where $\delta$ functions enforce energy and momentum conservation. $\A_{min}$ stands for the explicitly Minkowski part, which is calculable using the standard techniques. $\A_{cos}$ means cosmological correction. It is of order $\tau_i^{-2}/p^2\sim H^2/M^2$. $\rho$ may be viewed as a distribution (possibly complex) over $\Delta E=|\bm{p}_1|+|\bm{p}_2|-|\bm{p}_3|-|\bm{p}_4|$. We have to emphasize that we did not derive \eqref{assertion}, but rather assumed it based on observations in this subsection. This allows us to run positivity argument on $\A_{min}$ without worrying about the unknown $H^2/M^2$ correction since $\delta(0)$ dominates over $\rho(\Delta E=0)$ and thus one might use \eqref{the bound} and \eqref{the t bound} to extract part of the positivity information for an EFT in the cosmological background.

\subsection{Applying the bounds}\label{subsec:apply bound}

The tree level $2\to2$ amplitude corresponding to
\eqref{scattering Lagrangian} in the center of mass frame is
\begin{equation}\label{amplitude}
\A(s,t)=\left(-\frac{9}{4}\alpha_1^2-6\alpha_1\alpha_2-4\alpha_2^2+\frac{3}{2}\beta_1+2\beta_2\right)\frac{s^2}{M_p^2\Lambda^2}+2\beta_3\frac{s^2+t^2+u^2}{M_p^2\Lambda^2}+\frac{1}{2}\beta_4\frac{stu}{M_p^2\Lambda^4}.
\end{equation}
Detailed calculation of \eqref{amplitude} but with $\beta_4=0$ has been
presented in Ref.\cite{Baumann:2015nta}. The Lagrangian
\eqref{scattering Lagrangian} is actually non-relativistic, in
that uncontracted time derivatives are present. However, if
it indeed captures the physics of some UV-complete theory below the
high energy cutoff $\Lambda$ and above the decoupling scale
$E_{mix}\lesssim \epsilon_H^{1/2}H$ \cite{Cheung:2007st}, positivity derived from locality and unitarity should be
inherited. To the leading order in $H/\Lambda$, the positivity
bound \eqref{the bound} reads (without loss of generality we set
$G_{2X}=-1/2$ hereafter)
\begin{equation}\label{tree level bound}
\begin{aligned}
&(1+\epsilon_H)B_X+\epsilon G_{2XX}
\\&+\bigg[4B_X^2(-61-34\epsilon_H+34\epsilon^2_H-2\epsilon_H^{(1)})+4\epsilon (1+33\epsilon_H)B_XG_{2XX}\\&+2\sqrt{2}\epsilon^{1/2}(-23+5\epsilon_H)B_XG_{3X}+\epsilon(15+\epsilon_H)B_{XX}+4\epsilon^2(2G_{2XX}^2-G_{2XXX})\\&-2\epsilon G_{3X}^2+\sqrt{2}\epsilon^{3/2}(4G_{2XX}G_{3X}-3G_{3XX})\bigg]\hn^2\ge0.
\end{aligned}
\end{equation}
Applying bound \eqref{the t bound} yields ($n=1$)
\begin{equation}\label{tree level t bound}
\begin{aligned}
-B_X+4\bigg[\epsilon B_{XX}+(34-28\epsilon_H)B_X^2-10\epsilon B_X G_{2XX}-\sqrt{2}\epsilon^{1/2}B_X G_{3X}\bigg]\hn^2\ge0.
\end{aligned}
\end{equation}
We have to stress again that for \eqref{the bound} and \eqref{the t bound} to be trustworthy, all assumptions made in section-\ref{subsec:positivity in cosmology} have to hold.


In the calculation, we did not assume the correction is even
powers of $H/\Lambda$. But it turns out the leading order
correction is indeed $H^2/\Lambda^2$, as expected in the
introduction section, so the bounds do not distinguish between contraction and expansion. As commented before, the
cosmological evolution breaks time-translation symmetry of the
EFT, so it is not surprising that the bound in Minkowski spacetime
(or the limit $H/\Lambda\rightarrow 0$) may be violated for a relatively large
$H^2/\Lambda^2$, even if $H^2/\Lambda^2<1$.

\subsection{Discussion}\label{subsec:example}

\underline{\textit{In the limit $H/\Lambda\rightarrow 0$}}

We can ignore the $H^2/\Lambda^2$ correction. Without the corrections, the bounds \eqref{tree level
bound} and \eqref{tree level t bound} are
\begin{equation}\label{0 bound}
(1+\epsilon_H)B_X+\epsilon G_{2XX}\ge0,\qquad B_X\le0.
\end{equation}
For $1+\epsilon_H>0$, we have the positivity constraint on $B_X$
as
\begin{equation}\label{0 constraint}
\epsilon G_{2XX}\ge -(1+\epsilon_H)B_X\ge0.
\end{equation}
In comparison, the Minkowski bound only gives $G_{2XX}\ge0$ (see
Appendix-\ref{apdx:minkowski bound}). Ref.\cite{Melville:2019wyy}
reports $G_{4X}\le0$ (in the notation here, it corresponds to $B_X\le0$) for a shift-symmetric Horndeski Lagrangian
with only $G_2(X)$ and $G_4(X)$. The covariant theory ($c_T=1$)
studied here is essentially different from that in
\cite{Melville:2019wyy} ($c_T\neq 1$), in which the bound
$G_{4X}\le0$ is actually equivalent to the subluminal condition
$c_T\le1$. In fact, the $G_{4X}\le0$ bound in
\cite{Melville:2019wyy} is derived from $Y^{(2,1)}\ge0$ in
\cite{deRham:2017avq}. The second bound in Eq.\eqref{0 bound} has
similar origin to $Y^{(2,1)}\ge0$ and it also implies $B_X\le0$.
However, as $B_X$ also contributes in the tree level
$\pi\pi\to\pi\pi$ scattering, we have an additional bound in
Eq.\eqref{0 bound} which further constrains $B_X$. 

In addition, the subluminality of $c_s^2$ brings another bound
\begin{equation}\label{subluminal}
\sqrt{2}\epsilon^{1/2}G_{3X}\ge 4(1-\epsilon_H)B_X-2\epsilon G_{2XX}.
\end{equation}
This relation is automatically satisfied if $G_{3X}\ge0$ and
$\epsilon_H<1$. However, if $G_{3X}=0$ and $\epsilon_H>1$, it
requires $\epsilon G_{2XX}\ge -2(\epsilon_H-1)B_X$, which is
stronger than (\ref{0 constraint}) for $\epsilon_H>3$. Note it is possible to have $\epsilon_H\sim\mathcal{O}(1)$ if the scattering energy scale $M$ satisfies $H\sim E_{mix}\ll M\ll\Lambda$.

\underline{\textit{With $H^2/\Lambda^2$ correction}}

An interesting example is the slow-roll inflation, where
$0<\epsilon\simeq\epsilon_H\ll1$, so the bounds (\ref{tree level
bound}) and (\ref{tree level t bound}) with $H^2/\Lambda^2$
correction can be simplified.
To see this, setting $\epsilon=\epsilon_H=0$ in \eqref{tree level
bound} and \eqref{tree level t bound}, we have
\begin{equation}\label{infl bound}
B_X-244 B_X^2\hn^2\ge0,\qquad -B_X+136 B_X^2 \hn^2\ge0.
\end{equation}
This suggests $B_X=0$ unless we consider the slow-roll suppressed
parts. This implies that for the potential-driving dS inflation,
the $c_T=1$ beyond-Horndeski EFT \eqref{L_bH} reduces to GR.
However, for the $\dot \phi$-driving inflation, such as
k-inflation  \cite{ArmendarizPicon:1999rj,Garriga:1999vw} and
G-inflation \cite{Kobayashi:2010cm}, since $\epsilon\ll1$ may be
violated, the constraint on $B_X$ will be released.

\begin{figure}
\includegraphics[width=7in]{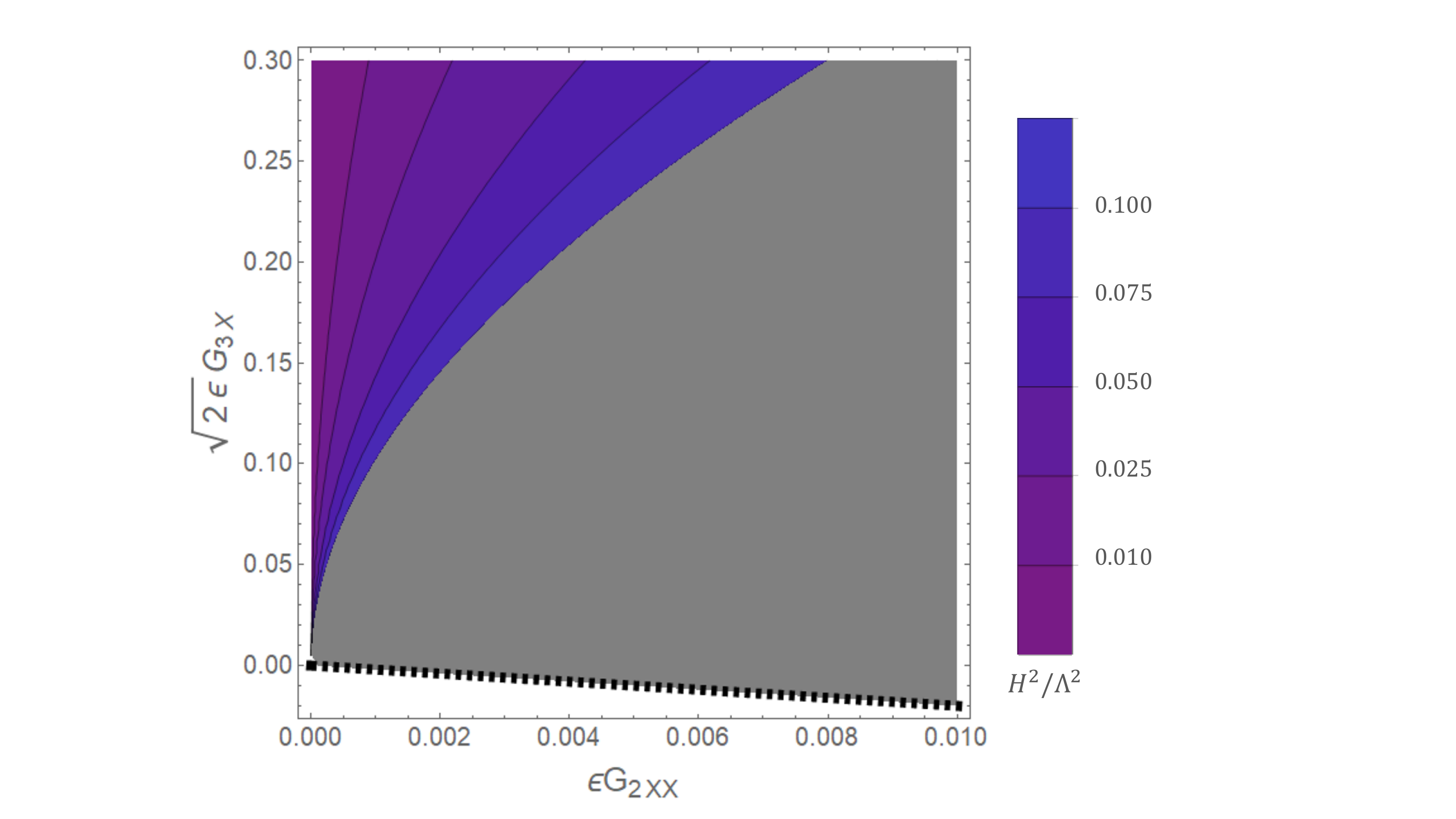}
\caption{The constrained phase space of $\epsilon G_{2XX}$ and $\sqrt{2\epsilon}G_{3X}$ by eq.\eqref{subluminal}, eq.\eqref{GR bound} together with the Minkowski bound $G_{2XX}\ge0$. The black dashed line is given by eq.\eqref{subluminal}. The curved lines are given by eq.\eqref{GR bound}. It can be seen that as cosmological evolution becomes important (i.e:$(H/\Lambda)^2$ gets larger), the phase space becomes more constrained for non-zero $G_{3X}$. Constraint lines with $(H/\Lambda)^2>0.1$ are not plotted since assumptions used to derive the bound break down when the scattering energy scale is comparable to the Hubble scale, hence the gray region in the plot.}
\label{phase}
\end{figure}

Inspired by the previous example, we look at the GR limit (we no longer assume $\epsilon\ll1$ here), in
which $\phi$ is not coupled to the Ricci scalar ($B\equiv1/2$).
The corresponding EFT is called Galileon \cite{Nicolis:2009qm}.
Assuming $G_{2XXX}=G_{3XX}=0$ for simplicity, we have
\begin{equation}\label{GR bound}
\epsilon G_{2XX}+\left[8(\epsilon G_{2XX})^2+4(\epsilon G_{2XX})(\sqrt{2}\epsilon^{1/2}G_{3X})-(\sqrt{2}\epsilon^{1/2}G_{3X})^2\right]\hn^2\ge0.
\end{equation}
The cosmological positivity bound itself might be weaker than its Minkowski counter part. For example, let $G_{3X}=0$ then cosmological bound \eqref{GR bound} also allows for a negative branch of $\epsilon G_{2XX}$ in addition to $\epsilon G_{2XX}\ge0$. We are also interested in the possibility of obtaining stronger bounds. The combined constraint of the cosmological bound \eqref{GR bound}, the subluminal condition \eqref{subluminal} and the Minkowski bound $G_{2XX}\ge0$ is plotted in Fig-\ref{phase} with a color coding for different values of $(H/\Lambda)^2$. The phase space of $\epsilon G_{2XX}$ and $\sqrt{2\epsilon}G_{3X}$ becomes more constrained as $(H/\Lambda)^2$ gets larger. Another point worth mentioning is one can now answer whether its UV completion would require the Minkowski bounds to get weaker/stronger when extended to cosmology. It is an interesting issue since future data/theoretical restrictions might be able to indicate whether the cosmological positivity bound is stronger or weaker than its flat space counterpart.

\section{Conclusion}\label{sec:conclusion}

We have investigated the application of positivity bounds in
the EFT of cosmological perturbations. As an illustrative example, we considered a
$c_T=1$ beyond-Horndeski EFT \eqref{L_bH}. We explicitly showed by
calculation that the leading cosmological correction to positivity
bounds indeed comes at $H^2/\Lambda^2$ and $\dot{H}/\Lambda^2$
order, consistent with our observation in the introduction. It is
also observed that the positivity bounds found in the limit
$H/\Lambda\rightarrow 0$ (or in Minkowski spacetime) might be violated
when $H$ is not far smaller than the cutoff scale $\Lambda$, since
the coefficient before $H^2/\Lambda^2$ correction is at $10^2$ order.
The bound with cosmological
corrections could be either weaker or stronger. We also discussed the
applications of our bound. It is found that positivity favors
a suppressed $B_X$ (comparable in size with $\epsilon G_{2XX}$ or
$\sqrt{2}\epsilon^{1/2}G_{3X}$) for slow-roll inflation.


Lagrangian \eqref{L_bH} can be used to implement fully stable
cosmological bounce. Nonpathological bouncing models built in
Ref.\cite{Cai:2017tku,Kolevatov:2017voe,Mironov:2018oec,Ye:2019frg,Ye:2019sth}\footnote{By "nonpathological" we require the bouncing models to be free of gradient and ghost instability throughout the \textit{whole} evolution history of the universe. In all the models cited, due to the higher derivative operators in theories beyond Horndeski, $c_s^2$ is always positive, thus there is no strong coupling problem and perturbative unitarity is preserved, see e.g.\cite{deRham:2017aoj}, see also \cite{Koehn:2015vvy} for showing the strong coupling problem in the $P(X)$ theory.} all have $G_{2XX}/G_{2X}<0$ somewhere, which seems to be inconsistent with the bound
(\ref{0 bound}) at first sight. There is no tension for now. Typically, to
violate the null energy condition (NEC, see \cite{Rubakov:2014jja}
for a review), one requires that the operator $X^2$ is not
negligible (depending on the value of $\phi$) at the NEC-violating
regime, while the beyond-Horndeski operator takes effect as a
stabiliser that controls gradient stability. Usually, such models
display obvious $\phi$-dependence and nonnegligible $\ddot{\phi}$
around the bounce point, which invalidates the assumptions we used
to derive the bounds here. Thus our bounds cannot be directly
applied to the bouncing models. It is possible to relax these
assumptions in more complete study. We might come back to
relevant issues in future works.

It is also interesting to integrate out the IR part of the RHS of
\eqref{branch int} within the regime of validity of the EFT to
give a more precise bound \cite{Bellazzini:2017fep}. Recently, the
positivity bounds with heavy spinning intermediate states have
been studied in inflation but from a covariant point of view
\cite{Kim:2019wjo}, and also Ref.\cite{Herrero-Valea:2019hde} has
explored positivity in the Higgs-Dilaton inflation model. It is
also well-motivated to go beyond the decoupling limit and study
the EFT with graviton and high-spin particles included
\cite{Arkani-Hamed:2015bza,Lee:2016vti}.

\textbf{Acknowledgments}

We would like to thank Brando Bellazzini, Matthew Lewandowski and Javi Serra for insightful comments and discussions on the t-channel singularity and positivity argument. GY would like to thank Yong Cai for helpful discussions. This work is supported
by NSFC, Nos.11575188, 11690021.

\appendix

\section{The Stuekelberg trick}\label{apdx:Stuekelberg}
In the unitary guage
\begin{equation}
X\equiv \partial_\mu \phi\partial^\mu\phi=\dot{\phi}^2 \tilde{g}^{00},
\end{equation}
where $\tilde{g}^{00}$ is given in Eq.\eqref{goldstone-metric}. To
expand Lagrangian \eqref{unitary L}, we still need the expression
of the extrinsic tensor. The normal of the uniform-$\phi$
hypersurface is
\begin{equation}\label{normal}
n_\mu=\frac{\partial_\mu \tilde{t}}{\sqrt{-\partial_\mu \tilde{t} g^{\mu\nu} \partial_\nu \tilde{t} }}=\frac{\delta^0_\mu+\partial_\mu\pi}{\sqrt{-\tilde{g}^{00}}}.
\end{equation}
Recall that we are allowed to raise and lower indices with the unperturbed FRW metric in the decoupling limit
\begin{equation}
n^\mu=\frac{-\delta^\mu_0+\partial^\mu \pi}{\sqrt{-\tilde{g}^{00}}}.
\end{equation}
The extrinsic curvature $K$ is
\begin{equation}
K=-\nabla_\mu n^\mu, \qquad K^\mu_{\ \nu} K^\nu_{\ \mu}=\nabla_\mu n^\nu \nabla_\nu n^\mu,
\end{equation}
with the relevant expressions as follows
\begin{subequations}\label{goldstone-extrinsic curvature}
    \begin{equation}
    \nabla_0 n^0=\partial_0\left(\frac{-1-\dot{\pi}}{\sqrt{-\tilde{g}^{00}}}\right),
    \end{equation}
    \begin{equation}
    \nabla_0 n^i=\partial_0\left(\frac{\partial_i\pi/a^2}{\sqrt{-\tilde{g}^{00}}}\right)+\frac{H\partial_i\pi/a^2}{\sqrt{-\tilde{g}^{00}}},
    \end{equation}
    \begin{equation}
    \nabla_i n^0=\partial_i\left(\frac{-1-\dot{\pi}}{\sqrt{-\tilde{g}^{00}}}\right)+\frac{H\partial_i\pi}{\sqrt{-\tilde{g}^{00}}},
    \end{equation}
    \begin{equation}
    \nabla_i n^j=\partial_i\left(\frac{\partial_j\pi/a^2}{\sqrt{-\tilde{g}^{00}}}\right)-\frac{(1+\dot{\pi})H\delta_i^j}{\sqrt{-\tilde{g}^{00}}}.
    \end{equation}
\end{subequations}
As a quick consistency check, one can use the above results to expand $K=-\nabla_\mu n^\mu$ to quadratic order in $\pi$
\[K=3H-\frac{\partial^2\pi}{a^2}+\frac{1}{2}H(\partial_i\pi)^2+\dot{\pi}\partial^2\pi+2\partial_i\dot{\pi}\partial_i\pi+\mathcal{O}(\pi^3).\]
It coincides with eq.(B.12) of Ref.\cite{Cusin:2017mzw} (neglecting metric fluctuations) except for the term $3H(t-\pi)=3H-3(\dot{H}\pi+\frac{1}{2}\ddot{H}\pi^2)+\mathcal{O}(\pi^3)$, an negligible mass term suppressed by $\dot{H}=-\epsilon_H H^2$.
According to the Gauss-Codazzi formula, the 3d Ricci scalar
$R^{(3)}$ is
\begin{equation}
R^{(3)}=R+2R_{\mu\nu}n^\mu n^\nu-K^2+K^\mu_{\ \nu} K^\nu_{\ \mu}.
\end{equation}
On the RHS, the Ricci scalar $R$ is invariant and $R_{\mu\nu}$ can be easily calculated using the spatially flat FRW metric
\begin{equation}
R_{00}=-3\frac{\ddot{a}}{a},\qquad R_{ij}=(a\ddot{a}+2\dot{a}^2)\delta_{ij}.
\end{equation}Now we are well-equipped to expand Lagrangian \eqref{unitary L} in powers of $\pi$ and its derivatives.

\section{Derivation of Goldstone Lagrangian \eqref{scattering Lagrangian}}\label{apdx:L calculation}
The equation of motion (EoM) of the free Goldstone field is
\begin{equation}\label{pi EoM}
\ddot{\pi}+3H\dot{\pi}-c_s^2\frac{\nabla^2\pi}{a^2}=0.
\end{equation}
Each $\pi$ field can have at most two derivatives. In fact, only
$K_{\mu\nu}$ contributes second derivatives of $\pi$. Thus the
n-th order Lagrangian $L^{(n)}$ contains at most $(n+2)$
derivatives. We first consider the part of $L^{(3)}$ with up to
four derivatives
\begin{equation}\label{general L3} g_1\dot{\pi}^3+g_2\dot{\pi}\frac{(\partial_i\pi)^2}{a^2}+g_3\dot{\pi}\frac{\partial_i\dot{\pi}\partial_i\pi}{a^2}+g_4\ddot{\pi}\frac{(\partial_i\pi)^2}{a^2}+g_5\dot{\pi}^2\frac{\nabla^2\pi}{a^2}+g_6\ddot{\pi}\frac{\dot{\pi}^2}{a^2},
\end{equation}
which includes all possible three point interactions with at most
four derivatives that may yield nonvanishing amplitudes in the
center of mass (CM) frame. Note that in the CM frame, vertices with no
time-derivatives (e.g:
$\partial_i\partial_j\pi\partial_i\pi\partial_j\pi$) do not
contribute in exchange diagrams since the exchanged virtue particle has vanishing
3-momentum. After some integration by parts, we have
\begin{equation}
S^{(3)}=M_p^2\Lambda^2\int dx^4 a^3\left\{g_1\dot{\pi}^3+(g_2-\don g_4)\dot{\pi}\frac{(\partial_i \pi)^2}{a^2}+\dot{\pi}^2\left[g_6\ddot{\pi}-(g_3/2-g_4-g_5)\frac{ \nabla^2\pi }{a^2}\right]\right\},
\end{equation}
where $\mathcal{D}_n\equiv nH+d/dt$ is defined, and $g_n, \
n=1,2,\dots,6$ are only dependent on time. Insert the EoM
\eqref{pi EoM} and switch to the rescaled coordinates and
normalised field, we get
\begin{equation}\label{L3}
\begin{aligned}
S^{(3)}=M_p^2\Lambda^2\int dx^4\sqrt{-g}&\left\{\left[g_1+\frac{g_2}{\cs}-\frac{\don g_4}{\cs}-\frac{1}{3}\dth g_6+(3H-\dth/3)\frac{-g_3/2+g_4+g_5}{\cs}\right]\dot{\pi}_c^3\right.\\&\left.+\left(-\frac{g_2}{\cs}+\frac{\don g_4}{\cs}\right)\dot{\pi}_c\frac{(\partial\pi_c)^2}{a^2}\right\}\frac{1}{(\sqrt{2c_s^3U}M_p\Lambda)^3}\\\equiv\int dx^4\sqrt{-g}&\left[\alpha_1\dot{\pi}_c^3+\alpha_2\dot{\pi}_c\frac{(\partial\pi_c)^2}{a^2}\right]\frac{1}{M_p\Lambda}.
\end{aligned}
\end{equation}
Now consider the part with higher-order derivatives
\begin{equation}
\frac{1}{a^4}\left[g_7\left(\dot{\pi} (\nabla^2\pi)^2-\dot{\pi}
(\partial_i\partial_j\pi)^2  \right)+g_8(\partial_i
\dot{\pi}\partial_i \pi \nabla^2\pi -\partial_i \dot{\pi}
\partial_j \pi \partial_i\partial_j\pi )\right].
\end{equation}
After some integration by parts, we find that the higher
derivatives cancel out and
\begin{equation}
-\frac{1}{a^4}(-H+\frac{d}{dt})\frac{g_8-g_7}{2} \nabla^2\pi ( \partial_j \pi)^2
\end{equation}
remains, which can be safely neglected in the CM frame. The
calculation of $S^{(4)}$ is similar but more involved. In
particular, higher-order derivative operators may contribute in
contact diagrams. We first consider operators with at most five
derivatives and then look at the higher derivative part. The part
of $S^{(4)}$ with at most five derivatives is
\begin{equation}
\begin{aligned}
&h_1\dot{\pi}^4+h_2\dot{\pi}^2\frac{(\partial_i \pi)^2}{a^2}+h_3\frac{(\partial_i \pi)^4}{a^4}+h_4\frac{ \partial_i\dot{\pi} \partial_i \pi( \partial_j \pi)^2}{a^4}+h_5\dot{\pi}\frac{ \partial_i\partial_j\pi \partial_i \pi \partial_j \pi}{a^4}\\&+h_6\dot{\pi}\frac{ \nabla^2\pi ( \partial_j \pi)^2}{a^4}+h_7\dot{\pi}\ddot{\pi}\frac{(\partial_i \pi)^2}{a^2}+h_8\dot{\pi}^2\frac{ \partial_i\dot{\pi} \partial_i \pi}{a^2}+h_9\dot{\pi}^3\frac{ \nabla^2\pi }{a^2}+h_{10}\ddot{\pi}(\dot{\pi})^3\\
\to&\left[h_1+(3H-\dth/4)\left(\frac{h_9}{\cs}-\frac{h_8}{3\cs}+\frac{h_7}{3\cs}+\frac{h_6}{3c_s^4}-\frac{h_5}{6c_s^4}\right)-\frac{\dth}{4}h_{10}\right]\dot{\pi}^4\\
&+\left[h_2+(\don/4-3H/2)\left(\frac{h_5}{c_s^2}-\frac{2h_6}{c_s^2}\right)-\frac{\don}{2}h_7\right]\dot{\pi}^2\frac{(\partial_i \pi)^2}{a^2}\\
&+\left[h_3-\frac{\mathcal{D}_{-1}}{8}(2h_4-h_5)\right]\frac{(\partial_i \pi)^4}{a^4}\\\equiv&A\dot{\pi}^4+B\dot{\pi}^2(\partial_i \pi)^2+C(\partial_i \pi)^4\\\to&\left[\left(A+\frac{B}{\cs}+\frac{C}{c_s^4}\right)\dot{\pi}_c^4-\left(\frac{B}{\cs}+\frac{2C}{c_s^4}\right)\dot{\pi}_c^2(\partial\pi_c)^2+\frac{C}{c_s^4}(\partial\pi_c)^4\right]\frac{1}{(\sqrt{2c_s^3U}M_p\Lambda)^4}\\
\equiv&\frac{1}{M_p^4\Lambda^4}\left[\beta_1\dot{\pi}_c^4+\beta_2\dot{\pi}_c^2(\partial\pi_c)^2+\beta_3(\partial\pi_c)^4\right].
\end{aligned}
\end{equation}

The highest derivative part of $S^{(4)}$ divides into to two
sectors. The first sector consists solely of spacial derivatives
\begin{equation}
\begin{aligned}
\sim &h_{11}\left[(\partial_k\pi)^2( \partial_i\partial_j\pi
)^2-(\partial_k\pi)^2( \nabla^2\pi )^2\right]+h_{12}(\partial_i
\pi \partial_i\partial_j\pi  \partial_j\partial_k\pi
\partial_k\pi-\partial_i\partial_j\pi\partial_i \pi \partial_j \pi
\nabla^2\pi ).
\end{aligned}
\end{equation}
It is easy to check that both the $h_{11}$ and $h_{12}$ vertices
cancel out in the CM frame. Vertices in the other sector are
\begin{equation}
\begin{aligned}
&h_{13}\left(\dot{\pi}^2 (\partial_i\partial_j\pi)^2-\dot{\pi}^2 (\nabla^2\pi) ^2\right)+h_{14}(\dot{\pi} \partial_i\partial_j\pi  \partial_i\dot{\pi}  \partial_j \pi-\dot{\pi} \partial_i\dot{\pi} \partial_i \pi \nabla^2\pi )\\
&+h_{15}\left( (\partial_i\dot{\pi} )^2 (\partial_j \pi)^2-\ddot{\pi}(\partial_i \pi)^2 \nabla^2\pi \right)+h_{16}\left( (\partial_i\dot{\pi} \partial_i \pi)^2 -\ddot{\pi} \partial_i\partial_j\pi \partial_i \pi \partial_j \pi\right)\\
\to&(h_{14}-2h_{13})(\dot{\pi} \partial_i\partial_j\pi  \partial_i\dot{\pi}  \partial_j \pi-\dot{\pi} \partial_i\dot{\pi} \partial_i \pi \nabla^2\pi )+h_{15}\left( (\partial_i\dot{\pi})^2 (\partial_j \pi)^2-\ddot{\pi}(\partial_i\pi)^2 \nabla^2\pi \right)\\&+h_{16}\left( (\partial_i\dot{\pi} \partial_i \pi)^2-\ddot{\pi} \partial_i\partial_j\pi \partial_i \pi \partial_j \pi\right).
\end{aligned}
\end{equation}
with the coefficients
\begin{equation}
\begin{aligned}
h_{13}&=6B+6\ndp^2B_X+4\ndp^4B_{XX},\\
h_{14}&=24B+16\ndp^2B_X,\qquad h_{15}=-h_{16}=4B.
\end{aligned}
\end{equation}
Performing integration by part again to the terms
proportional to $B(X)$, we have
\begin{equation}
\begin{aligned}
&\left[4\ndp^2B_X-8\ndp^4B_{XX}\right](\dot{\pi} \partial_i\partial_j\pi  \partial_i\dot{\pi}  \partial_j \pi-\dot{\pi} \partial_i\dot{\pi} \partial_i \pi \nabla^2\pi )\\&+\mathcal{D}_{-1}(6B) \dot{\pi} \nabla^2\pi  \partial_j \pi^2-\frac{B}{2}\frac{d^2}{dt^2}[(\partial_i \pi)^4].
\end{aligned}
\end{equation}
So in conclusion, the higher derivative operators introduce one new vertex $(\dot{\pi} \partial_i\partial_j\pi  \partial_i\dot{\pi}  \partial_j \pi-\dot{\pi} \partial_i\dot{\pi} \partial_i \pi \nabla^2\pi )$ and modify the coefficients
\begin{equation}\label{S4 modification}
h_3\to h_3-\frac{B}{2}(1+\epsilon_H)H^2,\qquad h_6\to h_6-6BH.
\end{equation}
Putting together all the results in this Appendix, we get
Lagrangian \eqref{scattering Lagrangian}.

\section{Coefficients in \eqref{scattering Lagrangian}}\label{apdx:coefficient}
We list here the explicit expressions of the coefficients in
\eqref{scattering Lagrangian}, calculated to the leading order in
$H/\Lambda$,
\begin{subequations}
    \begin{equation}
    \alpha_1=\frac{1}{\epsilon^{1/2}(-G_{2X})^{3/2}}\hn\left[(2+8\epsilon_H/3)B_X+\sqrt{2}\epsilon^{1/2}G_{3X}\right],
    \end{equation}
    \begin{equation}
    \alpha_2=\frac{1}{\epsilon^{1/2}(-G_{2X})^{3/2}}\hn\left[2(-1+\epsilon_H)B_X+\frac{1}{2}\sqrt{2}\epsilon^{1/2}G_{3X}+\epsilon G_{2XX}\right],
    \end{equation}
    \begin{equation}
    \begin{aligned}
    \beta_1=\frac{1}{\epsilon G_{2X}^2}&\left\{\frac{2 B_X (-3+\epsilon_H)}{3 }\right.\\&\left.+\frac{1}{3G_{2X}}\bigg[-12 B_X^2 \left(5 \epsilon_H^2-36 \epsilon_H+\epsilon^{(1)}_H+35\right)-14 B_X \epsilon (\epsilon_H-9) G_{2XX}\right.\\&\left.+3 \sqrt{2} B_X \epsilon^{1/2} (3 \epsilon_H+5) G_{3X}+B_{XX} \epsilon (\epsilon_H-27) G_{2X}\bigg]\hn^2\right\},
    \end{aligned}
    \end{equation}
    \begin{equation}
    \begin{aligned}
    \beta_2=\frac{1}{\epsilon G_{2X}^2}&\bigg\{B_X (3-\epsilon_H)\\&+\frac{1}{G_{2X}}\bigg[4 B_X^2 \left(10 \epsilon_H^2-53 \epsilon_H+\epsilon^{(1)}_H+49\right)+4 B_X \epsilon (4 \epsilon_H-15) G_{2XX}\\&-2 \sqrt{2} B_X \epsilon^{1/2} (\epsilon_H+2) G_{3X}\\&+\epsilon \left(12 B_{XX} G_{2X}-\epsilon G_{2X} G_{2XXXX}+2 \epsilon G_{2XX}^2+\sqrt{2} \epsilon^{1/2} G_{2XX} G_{3X}\right)\bigg]\hn^2\bigg\},
    \end{aligned}
    \end{equation}
    \begin{equation}
    \begin{aligned}
    \beta_3=\frac{1}{\epsilon G_{2X}^2} &\left\{\frac{B_X (3 \epsilon_H-5)+\epsilon G_{2XX}}{8}\right.\\&\left.-\frac{1}{8G_{2X}}\bigg[4 B_X^2 \left(42 \epsilon_H^2-121 \epsilon_H+85\right)+4 B_X \epsilon (29 \epsilon_H-42) G_{2XX}\right.\\&\left.+2 \sqrt{2} B_X \epsilon^{1/2} (3 \epsilon_H-5) G_{3X}+6 B_{XX} \epsilon G_{2X}+\sqrt{2} \epsilon^{3/2} (3 G_{2X} G_{3XX}+2 G_{2XX} G_{3X})\right.\\&\left.+20 \epsilon^2 G_{2XX}^2\bigg]\hn^2\right\}.
    \end{aligned}
    \end{equation}
    \begin{equation}
    \begin{aligned}
    \beta_4=&\frac{1}{\epsilon G_{2X}^2}\left[\frac{B_X}{2}\hn^{-2}+\frac{ (34-28\epsilon_H)B_X^2-B_X \left(10 \epsilon G_{2XX}+ \sqrt{2} \epsilon^{1/2} G_{3X}\right)-2 \epsilon B_{XX}G_{2X}}{ G_{2X}}\right]
    \end{aligned}
    \end{equation}
\end{subequations}

\section{Positivity in the Minkowski spacetime}\label{apdx:minkowski bound}
To derive positivity bounds in the Minkowski spacetime for
\eqref{L_bH}, one should switch back to $\dot{\phi}^2$ and
$\dot{H}$ by inserting \eqref{epsilon}, and pass to the
$H,\dot{H}\to0$ limit. The extrinsic curvature $K$ vanishes in the
Minkowski spacetime. Thus positivity only constrains $G_2(X)$
and its derivatives. The Minkowski bound as well as the sound
speed are ($G_{2X}=-1/2$)
\begin{equation}\label{minkowski bound}
\begin{aligned}
&G_{2XX}+\left(4 G_{2XX}^2-2 G_{2XXX}\right)\ndp^2\\&+ \left(40 G_{2XX}^3-4 G_{2XX} G_{2XXX}+\frac{G_{2XXXX}}{2}\right)\ndp^4+\dots\ge0,
\end{aligned}
\end{equation}
\begin{equation}
c_s^2=\frac{1}{1+4G_{2XX}\ndp^2}.
\end{equation}
Absence of superluminality places the bound $G_{2XX}\ge0$. An
interesting case is that if $G_2(X)\sim X^n$, in which $X^n$ is
the first nonnegligible higher-order derivative operator in the
$G_2(X)$ polynomial, the positivity bound \eqref{minkowski bound}
implies $G_{2XXX}\le0$ if $G_{2XX}=0$ and $G_{2XXXX}\ge0$ if
$G_{2XX}=G_{2XXX}=0$, consistent with the result of
Ref.\cite{Chandrasekaran:2018qmx}.

\end{document}